\documentclass[runningheads]{llncs}
\usepackage{graphicx}
\usepackage{mdframed}
\usepackage{paralist,tabularx}
\usepackage{enumitem}
\usepackage{footmisc}
\usepackage{amsfonts}
\usepackage{multirow}
\usepackage{subcaption}

\usepackage{amsmath}
\usepackage[export]{adjustbox}
\usepackage[user,titleref]{zref}
\DeclareMathOperator*{\argmax}{arg\,max}

\usepackage[T1]{fontenc}
\DeclareMathOperator*{\E}{\mathbb{E}}
\newenvironment{myfont}{\fontfamily{pcr}\selectfont}{\par}
\DeclareTextFontCommand{\textmyfont}{\myfont}

\usepackage{pifont}
\usepackage{xcolor}
\newcommand\blue[1]{%
  \bgroup
  \hskip0pt\color{blue}%
  #1%
  \egroup
}
\newcommand\red[1]{%
  \bgroup
  \hskip0pt\color{red}%
  #1%
  \egroup
}

\begin{document}
\title{Fast Few shot Self-attentive Semi-supervised\\ Political Inclination Prediction}

\titlerunning{Fast Few shot Self-attentive Semi-supervised Political Inclination Prediction}
%
\author{Souvic Chakraborty \and
Pawan Goyal \and
Animesh Mukherjee}
\authorrunning{Chakraborty et al.}

\institute{Indian Institute of Technology, Kharagpur, West Bengal, India \\
\email{\{chakra.souvic, animeshm, pawang\}@gmail.com}}

\maketitle              
\begin{abstract}

With the rising participation of the common mass in social media, it is increasingly common now for policymakers/journalists to create online polls on social media to understand the political leanings of people in specific locations. The caveat here is that only influential people can make such an online polling and reach out at a mass scale. Further, in such cases, the distribution of voters is not controllable and may be, in fact, biased. On the other hand,if we can interpret the publicly available data over social media to probe the political inclination of users, we will be able to have controllable insights about the survey population, keep the cost of survey low and also collect publicly available data without involving the concerned persons. Hence we introduce a self-attentive semi-supervised framework for political inclination detection to further that objective. The advantage of our model is that it neither needs huge training data nor does it need to store social network parameters. Nevertheless, it achieves an accuracy of 93.7\% with no annotated data; further, with only a few annotated examples per class it achieves competitive performance. 
We found that the model is highly efficient even in resource-constrained settings, and insights drawn from its predictions match the manual survey outcomes when applied to diverse real-life scenarios. 

\end{abstract}

\section{Introduction}
Political inclination refers to the political stance of an individual. 
Polling and surveying to understand the political leanings of people within a particular community, in a particular geopolitical region, or a specific context is a common approach. However, the manual polling mechanisms used today are hard to scale. Also, there is a significant chance of biased sampling as the samples are often too small in in terms of the number of individuals surveyed and localized. On the other hand, if a survey or poll is conducted on online social platforms, it is impossible to control voters' distribution to calibrate it to resemble a random sample of opinions. Often the voters in these polls are limited to being the active audience of the pollsters sharing similar political inclination. Thus the result of the same poll can be completely different if introduced by a different pollster. Therefore, algorithmic labeling of people chosen from a controllable distribution is important, rather than asking for bias-prone active participation by influencers sharing particular political inclination or cost-inefficient manual polling.

Most of the existing approaches of political inclination detection (PID) on social networks focus on probabilistic models~\cite{ijcai2017,predict_twitter,iyyer14,Sandeepa}, which are in turn based on the texts generated by users. Researchers have also tried to exploit the network structure by making use of GCNs~\cite{timme} (Graph Convolutional Networks). This method uses all second-degree features (neighbors of neighbors of the node/user to be classified in the graph) for a rich representation which makes the classification more accurate at the expense of speed. The data collection process involves collecting features of followers of followers of the user whose political inclination needs to be detected. As the collection of followers and all their tweets itself is a slow process limited by Twitter\footnote{\url{https://developer.twitter.com/en/docs/twitter-api/v1/rate-limits}}, the time required for the collection of the features of the second-degree neighbours increases quadratically in terms of average unique neighbors per node.

Also, the GCN-based models need to store the huge Twitter-subnetwork involving political persons and their followers. This severely violates the users' \textit{right to erasure} as per the Article 17 of GDPR\footref{gdpr} which reads as follows:


\begin{mdframed}[backgroundcolor=gray!10,font=\bfseries]\small``The data subject shall have the right to obtain from the controller the erasure of personal data concerning him or her without undue delay and the controller shall have the obligation to erase personal data without undue delay ..."
\end{mdframed}

Further, these models are trained on a huge number of annotated examples. This makes the approaches hard to scale for newer settings and countries. In contrast, we show that certain \textit{easy to collect features} plugged into a novel self-attentive framework can be very accurate in predicting the political inclination even if trained on a handful of annotated examples.

\noindent Our main contributions are as follows:\\
\noindent \textbf{(1)}. Graph-based methods used previously raise many ethical questions~\cite{koops2011forgetting,mat1,lindmeek,thompson2021ethical}. The users on social media platforms have the 
right\footnote{\url{https://gdpr-info.eu/art-17-gdpr/}\label{gdpr}}to deletion of their data from other storage systems which are dependent on social media as data source whenever their public profile on social media is deleted. Graph-based methods violate this by storing their information such as retweets, mentions, likes, and the follower-followee network. The time required to build and update such networks is huge as it will require everyday monitoring for (i) the existence of each connection and (ii) the arrival of new connections. So, the only way to use these features at inference time is to store them permanently in memory.
We eliminate the need for storing such large relational graphs from past social media data of a huge number of users. 
We achieve this by using richer first-degree features that we collect directly at inference time along with their second degree neighbors which can be collected from the tweets of the concerned user/person to be classified directly (e.g., we collect the hashtags used by the retweeted user as it is readily available with the retweet, same for replies). Using smart augmentation of these features we beat the performance obtained by the graph-based approaches~\cite{timme} at a reduced inference time. 

\begin{figure}
    \centering
    \includegraphics[width=0.45\textwidth]{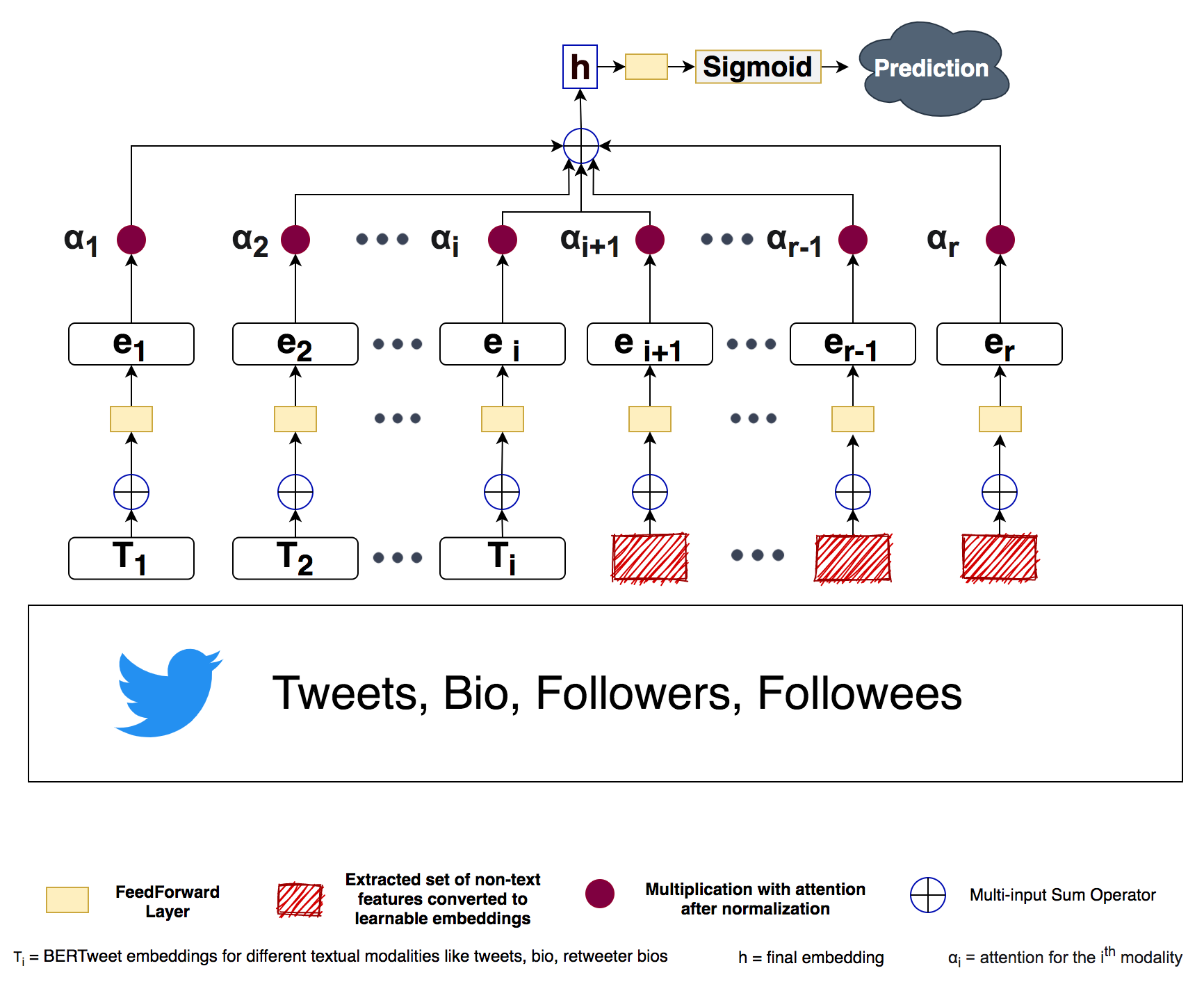}
    \caption{\scriptsize{The FSSPIP base architecture: The Twitter profile of a user is taken as input from which 22 different feature types are extracted and processed to predict political inclination. 
    }}
    \label{fig:arch}
\end{figure}

\noindent \textbf{(2)}.  We propose a novel Fast Self-attentive Semi-supervised Political Inclination Predictor \emph{FSSPIP} (Figure~\ref{fig:arch}). The experimental results show that even without using any gold annotation, we can achieve high accuracy of $\sim$ 94\% using weak supervision. The model is highly scalable and free from manual intervention unlike Darwish et al (2020)\cite{UUS} which needs human supervision or cluster inspection. 


\noindent \textbf{(3)}. We bring on board multiple additional datasets to show that our model can be used in many other similar settings for political inclination detection with a handful of labeled examples (or even without it). In specific, we present several case studies on media bias and political polarization using our classifier in zero-shot settings. 

\vspace{-0.2cm}
\section{Related work}
\vspace{-0.2cm}

Stefanov et al (2020)\cite{stefanov} and Baly et al(2020) \cite{baly2020} used Wikipedia, Twitter, YouTube, and other channels of information to detect the political leanings of the media houses. This approach is not scalable in context of persons.  Conover et al(2011)\cite{predict_twitter} had used a corpus of 1000 annotated data points to test the supervised approaches based on bag of words. 
Iyyer et al(2014)\cite{iyyer14} used advanced neural techniques like RNNs on a labeled corpus of sentences taken from speeches of democratic and republican parliamentarians. Chen et al(2017)\cite{ijcai2017} used graph-based approaches to show the efficacy of using an opinion-aware knowledge graph. However, these techniques fail to take the richer network features into account. They also completely rely on annotated data failing to take advantage of domain knowledge of the task in hand. 


Aldayel et al (2019) \cite{NTF} analyzed the features responsible for higher accuracy in stance detection setups using  network features, tweet texts and text derived features.
Darwish et al (2020)\cite{UUS} on the other hand used clustering based unsupervised setup to detect stance of users mainly relying on three channels of features: retweeted tweets, retweeted accounts and hashtags.
Xiao et al (2020)\cite{timme} approached the same task using manual annotation and collecting a large dataset of non-politician social media users and politicians on Twitter. They relied on variants of relational GNNs coupled with multi-task learning. However, given the need for explicit storage of information in the graph structures, even after the training phase, the graph-based algorithms often violate privacy rights of a large section of users. 

Therefore, in this paper, we attempt to solve political inclination detection in a resource-constrained setup with no storage of user data after model training. 
We use several task-dependent augmentation techniques and unsupervised learning methods which have not been used in this context earlier thus making our model robust, easily adaptable and scalable without any human help/supervision. 
We only use public data available at the time of inference. 



\vspace{-0.2cm}
\section{Model architecture}
\zlabel{model} 

\vspace{-0.2cm}
\noindent\textbf{The Base Architecture}:
Like previous  state-of-the-art approaches~\cite{timme} we too use a \textit{GCN-like framework}. However, we, in contrast, do not store the user/feature graphs nor do we need a list of politicians in the country of the users to be classified with huge set of labelled examples. We use a long list of different \textit{feature types} derived from \textit{follow}, \textit{mention}, \textit{reply}, \textit{retweet}, \textit{tweets} and \textit{likes}. We hypothesize that for a good representation of the political inclination, there are many important but easy to collect features which can be retrieved from the web directly during inference time with no need of storage. 
We describe these features in details below:

\noindent\textbf{Base features}:

\noindent \textit{User descriptions}: 
We collected user descriptions of users retweeted and quoted, forming two separate documents. These user descriptions/bios often contain key information like the user's occupation, gender, religion etc. 

\noindent \textit{Hashtags}: Hashtags are important as similar hashtags are used to express opinions for/against a polarizing topic by users of different leanings. 

\noindent \textit{Mentions}: IDs mentioned in tweets are used as features.

\noindent \textit{Media domains}: It is no secret that users of different political leanings share different sets of news items that fit their ideological perspective. Considering their importance in our task, we collect domain names and domain + co-domain names from users' tweets. We use them as separate features.  

\noindent \textit{Textual content}: 
We use pre-trained models like BERTweet~\cite{bertweet}, and Google's Universal Sentence Encoder \cite{googuni} to convert the content of tweets of a user into embeddings. In our experiments, we found that BERTweet performs better (possibly because BERTweet is trained on text with vocabulary more similar to ours). Thus we report BERTweet numbers only.

\noindent \textit{Features connected to neighborhood}\label{extrafeat}: 

We mentioned a total of 6 features till now. We repeat the same features for retweets and replies separately. So, the total number of features become 6+6+6=18.

In addition to all these features, we also use \textit{friend ids}, \textit{follower ids}, \textit{mention ids}, \textit{ids replied to}, and \textit{ids retweeted} as features collected at test time. 

So, in total we use 18+4=22 features.

\noindent\textbf{Attending to different modalities}: 

We use $|R|=22$ features in our architecture. For each feature type $r$ and user $i$, we obtain an embedding $e_{ir}$ of size $d=8$ as follows.

\begin{equation} \label{eq1}
\footnotesize
\begin{split}
    e_{ir} & = W_r \times BERTweet(T_{ir}) ,\ \textrm{if} \ r \in T \\
    & = W_r A_{ir} H_r ,\ if \ r \in T' \\
\end{split}
\end{equation}

where $A_{ir} \in \{0,1\}^{1 \times Vlen_r}$ is the feature presence-absence vector for the $r\textsuperscript{th}$ feature and $H_r \in \mathbb{R}^{Vlen_r \times d}$ is the embedding matrix containing embeddings of all the features for feature type $r$. While pre-processing we chose only those features in the vocabulary which appeared at least in five instances of the training data to ensure having enough training instances. The length of the vocabulary for $r\textsuperscript{th}$ feature type is represented as $Vlen_r$. $T'$ and $T$ are respectively the set of non-textual and textual feature types. For each feature type $r$, $T_r$ denotes the textual content of that feature for user $i$, $W_r \in \mathbb{R}^{d \times d_{em}}$, where $d_{em}$ is the embedding dimension of the output of BERTweet~\cite{bertweet}.
Now, we calculate $h_i$, the final embedding for the $i^\textrm{th}$ user as follows.
\begin{equation} \label{eq2}\footnotesize
h_i=\sum_r \alpha_{ir} \times \frac{e_{ir}}{|e_{ir}|}
\end{equation}


FSSPIP uses a dynamic dot-product self-attention mechanism to calculate the weights for each of the feature types to finally get a weighted sum of the normalized embeddings of each feature type. We use learnable parameters $p,q,k \in [0,1]$ to allow some flexibility in attention calculation. Learnable parameters $q_r$ and $k_r \in R^d$ are queries and keys, respectively, for each feature type $r$ (Here a feature type is specific social media attribute, so a collection of hashtags coming from tweets is a feature type different from the collection of hashtags coming from the retweets/replies.Please refer to the list of features mentioned at the start of the section for a broader understanding). So,

\begin{equation}\label{eq5}\footnotesize
    \alpha_{ir} =p*\frac{e^{q_{ir} \times k_{ir}}}{\sum_r e^{q_{ir} \times k_{ir}}}+(1-p)*|e_{ir}|
\end{equation}
\begin{equation}
    q_{ir} =q \times e_{ir}+(1-q)\times q_r 
\end{equation}
\begin{equation}
k_{ir} =k \times e_{ir}+(1-k) \times k_r
\end{equation}
An illustration of this base architecture is presented in Figure~\ref{fig:arch}. It shows how input from each feature type goes through different transformation functions (BERTweet in case of textual data, trainable embeddings in case of follower ids etc.) to transform into embeddings which are then weighted by attention values calculated through three different architectural scheme as mentioned. The weighted summation of the embeddings (vector size 768) denote representation of the node/person to be classified. This embedding is further multiplied with a vector of size 768x1 and passed through a sigmoid function to obtain probability of a person being a republican in a binary classification setup. We use binary cross entropy as the loss function for supervision.
\vspace{-0.3cm}
\section{Augmented Semi-supervision for Superior Representation Learning}\zlabel{few-shot}
\vspace{-0.3cm}
To make our architecture ready for few-shot learning, we make the model robust using regularization and multi task learning. We also use a specific kind of weak supervision needing no human intervention producing high accuracy without any labelled example. Specifically, we use three different categories of techniques which are described below.


\noindent\textbf{Dynamic augmentation}: 

\noindent \textit{Mixup}: Mixup~\cite{zhang2018mixup} is a technique that enforces linear change in output given linear change in input by training a neural network on convex combinations of pairs of examples and their annotated labels for a particular task. We adopt the method to our network data by mixing two random users for each channel (e.g. : hashtags, domains, retweetees from both users are present for the augmented user) increasing the diversity of data points and regularizing the model for unseen data points.\\
\noindent \textit{Sampling}: Twitter users can be imagined as generative agents who generate tweets on selective issues and follow/reply/mention/interact with other users following some implicit probability distribution. So, if some of the points from the distribution are sampled out uniformly, the distribution will not change. So, we uniformly(chosen from a random uniform distribution for each feature type) sample out features from labeled examples for augmentation masking out 0-15\% of the features randomly during training.\\
\noindent \textit{Feature channel dropout}: While some feature types may influence the result more than others, it is important to learn to predict from the cues available if one influential feature type (e.g. \textit{hashtags}, \textit{follower}s, \textit{retweets} etc.) is absent. So, we randomly drop random feature types while training for better performance through adversarial training.\\
\noindent\textbf{Weak supervision}: 
We hypothesize that the followers/retweeters of a particular political party often share the bias of having that particular political inclination. So, they are statistically more likely to follow the leaning of that particular political party which they are following on social media than any other. This provides some silver labels in the Twitter space for weak supervision. We crawled the Twitter handles of each political party (i.e., the official Twitter handles of \textit{The Democratic} \& \textit{The Republican} party in case of US and \textit{AAP}, \textit{Congress} \& \textit{BJP} party in case of India) to collect the last 75,000(set heuristically to contain enough examples) followers and the last 75,000 retweeters for each of these parties. We randomly selected 2,500 from each of them to get a sample representative of the timeline (as the most recent followers appear first, so collecting a big pool and resampling may help) and collected their relevant data for training. Users following both parties were removed. \\
\noindent\textbf{Self-supervision}: Self-supervision is a semi-supervised learning technique that trains the model in a new dummy task predicting part of input data using the other part of the data~\cite{bertweet}. While in case of textual data masked language modelling and next sentence predictions \cite{bertweet,devlin-etal-2019-bert} are the most frequently used pre-training technique, graph neural nets use predicting masked edges between nodes as the pre-training task. Following these methods, we pretrain our model with the task of prediction of the non-textual features which are masked in the dynamic augmentation phase during sampling the input features. We use self-supervision as pre-training method while performing few-shot learning and fine-tune later on the annotated data points. Hyperparameter details and loss function of the pretraining phase has been put in the Appendix \ref{appendix:new}).
\vspace{-0.2cm}

\section{Data preparation}
\noindent \textbf{Dataset for the main task:} As provided by Xiao et al(2020)\cite{timme}, we have 2,976 labeled data points (labelled \textit{republican} or \textit{democratic}) along with 583 politicians' data in the US setting. For a nuanced analysis, we retain the partition of the data points, used in the dataset -- \textbf{PureP}\footnote{This dataset contains only the politicians.}, \textbf{P50}\footnote{This dataset contains people highly interested in politics being followed or following at least 50 politicians, including the politicians themselves.}, \textbf{P20$\sim$50}\footnote{This dataset contains people moderately interested in politics being followed or following anywhere between 20 to 50 politicians and the politicians themselves.} and \textbf{P+all}\footnote{This dataset contains members of \textbf{PureP}, \textbf{P20$\sim$50} along with many outliers who are following or being followed by maximum five politicians.}.

\begin{table}[ht]
\centering
\scriptsize
\begin{tabular}{lllll}
\hline
                     & PureP & P50   & P20$\sim$50 & P+all  \\ \hline
\#User               & 579   & 730   & 946         & 1,125    \\ \hline
\#Retweetees         & 26,034   & 46,734  & 67,239        & 87,802    \\ \hline
\#Repliees           & 1,738  & 16,261 & 12,665       & 29,713 \\ \hline
\#Mentions           & 30,363   &  53,125 & 62,274       & 84,962  \\ \hline
\end{tabular}%
\caption{\small{Descriptive statistics of the labeled dataset.}}
\label{tab:stat}
\vspace{-8mm}
\end{table} 

We collected the Twitter ids and labels provided by Xiao et al (2020)\cite{timme}. We crawled the last 3,200 tweets (some tweets got deleted, some tweets were retweets, quotes, and replies), follower ids and friend ids of each labeled id in November, 2020 using the Twitter API\footnote{\url{https://developer.twitter.com/en/docs/twitter-api}\label{f111}}. We also collected the user objects (containing bios) for each id. So, after pre-processing, we have data for each feature type described in the previous section. We extracted the domain and co-domain names from the URLs shared using the tldextract\footnote{\url{https://pypi.org/project/tldextract/}} library. Out of 2,976 labeled users, 2,665 users were available on Twitter at the time of crawling of the tweets (November, 2020). We report our results on this dataset. A major point to be noted here is that we do not store this data once the training is over, nor do we need to collect neighborhood data at inference time making the inferece faster and memory efficient. 
A detailed statistics of this dataset with the count of unique features for some feature types is provided in Table~\ref{tab:stat}. 


\subsection*{Additional datasets for lateral verification}
We collect several other datasets to demonstrate the usefulness of FSSPIP in zero/few-shot setting.
The statistics of these datasets are detailed in Table~\ref{tab:mediabias}.

\begin{table}[ht]
\centering
\scriptsize
\begin{tabular}{lllllll}
\hline
                     & MB& C & MP& S & TPC & HTU  \\ \hline
\#User               & 806      & 400       & 1000       & 2900      & 4000      & 4000    \\ \hline
\#Domains            & 5,629    & 3,829    & 6,651    & 17,285    & 25,679    & 27,287  \\ \hline
\#Hashtags           & 83,245   & 18,109    & 28,891   & 69,341   & 71,189   & 68,423  \\ \hline
\#Mentions           & 52,178 &96,123    & 67,478   & 89,651   & 89,765   & 81,319  \\ \hline
\end{tabular}%
\caption{\small{Descriptive statistics of the collected datasets. MB: MediaBias; C: Community; MP: Multiparty; S: Statewise TPC: Topicwise; HTU: HashTagUsers (4 hashtags subset as mentioned in Figure~\ref{fig:hashtag_india2} , details in Appendix \ref{appendix:hashindia}).}}
\label{tab:mediabias}
\vspace{-8mm}
\end{table}

\noindent\textbf{The media bias dataset}: Following Stefanov et al (2020)\cite{stefanov}, we use the crowdsourced labels\footnote{\url{https://mediabiasfactcheck.com/}} for media bias prediction. There are 806 labeled instances in the dataset with labels \textit{left, center-left, least biased, center-right} and \textit{right}. In order to binarize the label space (to fit in our classification model which is a binary classifier), we first discard the instances with label \textit{least biased}; next, we merge \textit{left, center-left} to a single label \textit{left} and \textit{center-right, right} to a single label \textit{right}. We collect the friend ids, follower ids, and the last 3,200 tweets of these media houses to employ the FSSPIP classifier for prediction.

\noindent\textbf{The ethnic community dataset}: Many post-poll surveys establish how different communities vote differently. We try to use our model to identify such divisions. We first sample recent tweets using the Twitter API\footref{f111} mentioning names of any of the communities. Among the users tweeting, we select only those who mention their community as one (or more) of the communities/ethnicities being probed (`black', `white', `hispanic/latino', `asian'), in their bio. We put a user to a particular community if that community is mentioned in their bio.

\noindent\textbf{Multi-party leaning dataset}: In order to collect a set of users residing in a multi-party democratic system, we filter the latest 10,000 tweets (and tweet-ers) containing the term `Delhi Election' using Twitter API on 17\textsuperscript{th} March, 2021. We annotate random 1000 Twitter users from this list into followers of three political parties: \textit{AAP: Aam Aadmi Party}, \textit{BJP: Bhartiya Janta Party} and \textit{Congress/INC: Indian National Congress} (AAP: 203 users, INC: 435 users, BJP: 362 users) to form the multi-party inclination dataset. We check the residency of the users and confirm it to be India through the self-declared location tag in twitter while annotating each user.

\noindent\textbf{Statewise inclination dataset}: Here, we use the Twitter API to collect tweets against a politically neutral query `election' (all datapoints are collected before 17.05.2021). If the user has a state's name mentioned in the \textit{Twitter's location tag}, we categorise that user to that particular state. We collected 100 users for each state in India for representative sample collection.

\noindent\textbf{Hashtags user data}: In order to find out the inclination distribution behind each hashtag, we collect the tweets containing some trending hashtags (on or before 17.05.2021). We collect 1000 (x 30) tweets, excluding the retweets and replies containing each trending hashtags using the Twitter API. 
For a manual verification, we annotate 30 hashtags with tags Congress, BJP, and Neutral (In the date of collection, we could not find hashtags which can be attributed to be inclined towards AAP. Moreover, politically unmotivated hashtags are termed as `neutral'). This annotation was done by a PhD student, expert in Indian Politics by reading the tweets with the hashtags.

\vspace{-2mm}
\section{Main task: Experiments and analysis}

\noindent\textbf{Baselines}: 
We use the best performing models provided by Aldayel et al (2019) \cite{NTF} and Darwish et al (2020)\cite{UUS} (UMAP+DBSCAN with tweets containing chosen hashtags included in the Appendix \ref{appendix:new}). \textbf{NTF} \cite{NTF} uses network/graph and textual features together in its model just like our model without attention. \textbf{UUS} \cite{UUS} on the other hand uses weak supervision (a quite different method compared to ours) through dimensionality reduction and clustering, manual inspection (which also makes the algorithm less scalable) and labelling of the clusters with only three features (retweeted tweets, retweeted accounts and hashtags). We also added a modified version of the UUS algorithm for a fair comparison with our fine-tuned model as the UUS algorithm is completely unsupervised and incapable of using any supervisory signal for few-shot learning. We took the unsupervised UUS model and fine-tuned it using annotated data points, terming it \textbf{UUS+}.

We also add non-neural baselines like SVM, Logistic regression (LR) and Random Forest (RF) as we are interested to show how simple algorithms with smaller inference time tally with our methods. Here, we use the concatenation of tweets for each user as input. We used TIMME-hierarchial~\cite{timme} and its other two variants as the other baselines using self-supervision on graphs with higher inference time due to second order data collection on large graph. However, we only report TIMME-hier results as it was the best-performing variant (hyperparameter stats and details on other TIMME variants in Appendix \ref{appendix:performance}). A qualitative comparison of the baselines is added in Table~\ref{tab:compare}.

\begin{table}[ht]
\centering
\scriptsize
\begin{tabular}{l|l|l|l|l|l|l}
\hline
Models                                               & FSSPIP          & TIMME     & NTF       & UUS+          & UUS       & NNeur                       \\ \hline
Uses only neighbors?                                 &\ding{51}             &           &\ding{51}  &\ding{51}      &\ding{51}  & \ding{51}                        \\ \hline
Uses textual features?                               &\ding{51}             &\ding{51}  &\ding{51}  & \ding{51}              &  \ding{51}         & \ding{51}                         \\ \hline
\begin{tabular}{@{}c@{}c@{}}\tiny{Uses unsupervised pretraining} \\ also while finetuning \\ on labelled examples?\end{tabular}                       &\ding{51}             &\ding{51}  &           &  \ding{51}    &           &                       \\ \hline
\begin{tabular}{@{}c@{}}Performs well even \\ 
without manual annotation?\end{tabular}                      &\ding{51}             &           &           &  \ding{51}    & \ding{51} &                      \\ \hline
\begin{tabular}{@{}c@{}}Performs well even without \\ 
any human supervision ?\end{tabular}                 &\ding{51}             &           &           &               &           &                      \\ \hline
\end{tabular}
\caption{\scriptsize{A pointwise comparison of the models used as baselines. \{NNeur : Non Neural baselines\}. }}
\label{tab:compare}
\end{table}
\vspace{-8mm}

\begin{table}[!t]
\centering
\fontsize{7pt}{8pt}\selectfont

\begin{tabular}{l|l|lll|llll|l}
\hline
         Dataset        & \#T & LR  & RF    & SVM    & NTF  & UUS  & UUS+ & TIMME&  FSSPIP      \\ \hline
\multirow{3}{*}{PureP}  & 50  & 81.8 & 90.1 & 79.2   & 89.6 & 94.1 & 95.2 &95.4  &  \textbf{97.1}$^{\dag}$       \\
                        & 250  & 85.1 & 93.2 & 83.8  & 93.2 & 94.1 & 97.5 &95.8  & \textbf{98.9}$^{\dag}$    \\
                        & 500 & 89.2 & 96.9 & 88.5   & 97.9 & 94.1 & 98.1 &99.2  & \textbf{99.2}    \\
                        \hline
\multirow{3}{*}{P50}    & 50   & 78.2 & 81.7 & 64.2  & 90.9 & 92.2 & 95.5  &84.6  & \textbf{97.3} $^{\dag}$   \\
                        & 250  & 79.2 & 83.8 & 65.4  & 92.3 & 92.2 & 96.8 &95.1  &  \textbf{98.1}$^{\dag}$ \\
                        & 500 & 80.4 & 84.1 & 69.3   & 97.8 & 92.2 & 97.5 &95.2  &  \textbf{98.9} $^{\dag}$  \\
                        \hline
\multirow{3}{*}{P20-50} & 50   & 86.1 & 86.6 & 81.2  & 90.2 & 92.5 & 94.9 &79.1  & \textbf{96.9} $^{\dag}$    \\
                        & 250  & 89.6 & 87.3 & 84.8  & 91.8 & 92.5 & 95.8 &96.9  &   \textbf{97.8}$^{\dag}$    \\
                        & 500 & 92.1 & 89.9 & 86.1   & 97.7 & 92.5 & 97.1 & 97.1  &  \textbf{98.9}  \\
                        \hline
\multirow{3}{*}{P+all} & 50   & 84.2 & 77.1 & 76.2   & 91.5 & 89.1 & 92.8 &69.5  & \textbf{95.2} $^{\dag}$  \\
                        & 250  & 86.5 & 79.5 & 77.9  & 93.9 & 89.1 & 94.5 &93.8 &  \textbf{97.1}$^{\dag}$  \\
                        & 500 & 87.6 & 83.2 & 80.8   & 96.7 & 89.1 & 97.4 &95.2  &   \textbf{98.1}$^{\dag}$  \\
                        \hline
TTI (secs)            &      & 28.0   & 28.9   & 28.1& 65.3 & 63.1 & 63.2  &973.2  & 62.1  \\ \hline 

\end{tabular}%

\caption{\scriptsize{Results of few-shot learning \{\textrm{NTF}: \textrm{Model proposed by \cite{NTF}}; \textrm{UUS}: \textrm{Model proposed by \cite{UUS}}; \textrm{UUS+}: \textrm{Model proposed by \cite{UUS} fine-tuned on annotated data points}; \textrm{TIMME}: \textrm{TIMME-hier (other TIMME variants' result in Appendix \ref{appendix:performance})}; \textbf{\textrm{FSSPIP}: \textrm{FSSPIP base architecture with the few-shot learning framework}};  \textrm{\#T}: \textrm{Number of training datapoints}; \textrm{TTI}: \textrm{Time Taken for Inference per datapoint with Twitter ids as inputs. For each framework, it also includes the time taken to collect the data}\}.}}
\label{tab:fewshot}
\vspace{-8mm}
\end{table}

\noindent\textbf{Results}: In Table~\ref{tab:fewshot}, we show that our best performing model \textit{FSSPIP}\footnote{$\dag$ represents p value less than 0.05 in student's t-test while comparing FSSPIP's result with the best performing baseline.} fairly beats other baselines for all datasets. We see that we gain most compared to other models when very few training datapoints (50) are present\footnote{Training on the whole data, we got comparable accuracy with other state-of-the-art architectures (result in Appendix \ref{appendix:performance}).}. In the case of the non-politician datasets, i.e., \textbf{P50, P20-50} and \textbf{P+all}, the performance obtained by our model is significantly higher than other baselines, even with only 50 training data points. This may be because the non-politician datasets do not purely contain political features unlike the PureP dataset making the feature learning task less straight-forward needing finer features like domain names a user is interested in or the tweets from the retweetees.

Our model performs better than other models in terms of time required to predict for a single user. Compared to the networks using second order relational data (TIMME) we are at least $\sim$ 10x faster as shown in Table~\ref{tab:fewshot}.

Also, our model performs better than NTF~\cite{NTF} and UUS~\cite{UUS} by a significant margin using weak supervision\footnote{We do not use any human intervention, unlike other approaches which cluster the datapoints and identify the cluster's political affiliation by manually sampling users and annotating them. This may also be subjected to randomness in the clustering process and dependent on the characteristics of the specific subsets of the social network.} with better augmentation while utilizing carefully extracted network features like NTF inputs~\cite{NTF}.

\noindent\textbf{Ablation study}:

\begin{table}[t]
\centering
\fontsize{8pt}{8.5pt}\selectfont

\begin{tabular}{l|l|llllll}

\hline
         Dataset                & \#T     & F1    &F2     & FSSPIP- - - & FSSPIP- -  & FSSPIP-   & FSSPIP \\ \hline
\multirow{3}{*}{PureP}          & 50      & 90.4  &91.3   & 91.6  & 92.3 & 94.2  & 97.1          \\ 
                                &250      & 94.1  &94.7   & 94.8  & 95.1 & 98.1  & 98.9 \\ 
                                & 500     & 98.9  &98.7   & 99.0$^{\dag}$  & 99.1 & 99.1  & 99.2 \\ \hline
\multirow{3}{*}{P50}            & 50      & 87.2  &89.4   & 90.9$^{\dag}$  & 92.1 & 94.9  & 97.3      \\ 
                                & 250     & 92.1 &91.5    & 92.7$^{\dag}$  & 93.5 &97.3   & 98.1   \\ 
                                & 500     & 95.6 &96.8    & 98.4$^{\dag}$  &98.9  & 98.9  & 98.9 \\            
                        \hline
\multirow{3}{*}{P20-50}         & 50     & 87.1 & 87.6  &  90.2$^{\dag}$    &92.8 & 93.8   & 96.9         \\ 
                                & 250    & 91.6 &92.4  &  92.8     &95.2 & 97.2   & 97.8     \\ 
                                & 500     & 97.8 &98.2 &  98.4     &98.5 & 98.5   & 98.9      \\ 
                        \hline
\multirow{3}{*}{P+all}          & 50     & 88.3 &88.5  &  89.2$^{\dag}$     &92.5 & 93.0   & 95.2       \\ 
                                & 250    & 92.1 &93.8  &  94.3$^{\dag}$     &95.8 & 96.9   & 97.1      \\ 
                                & 500    & 95.9 &96.3  &  97.8$^{\dag}$     &97.8 & 97.9   & 98.1       \\ 
                        \hline
\end{tabular}%

\caption{\scriptsize{Ablation study of different model variants \{\textrm{F1}: \textrm{FSSPIP-fixedattn}; \textrm{F2}: \textrm{FSSPIP-auto};  \textrm{FSSPIP- - -}: \textrm{FSSPIP base architecture}; \textrm{FSSPIP- -}: \textrm{FSSPIP without weak supervision and self supervision}; \textrm{FSSPIP-}: \textrm{FSSPIP without self supervision}.\}}}
\label{tab:ablation}
\vspace{-8mm}
\end{table}

\noindent\textit{Model variants} - 
In order to ablate our attention mechanism we employ two other varieties of attention in place of ours in FSSPIP base architecture which are as follows:
We recall equation~\ref{eq2} here to understand the two new attention mechanisms:
\noindent \textit{FSSPIP-fixedattn} (F1): FSSPIP-fixedattn uses fixed learnable attention to calculate a weighted sum of embeddings of each feature type.
    Thus here the equation~\ref{eq2} $\alpha_{r}$ values are learnable parameters and $\alpha_{ir}=\alpha_{r}$, $\forall i$.

\noindent \textit{FSSPIP-auto} (F2): FSSPIP-auto simply sums up each of the normalized embeddings of each feature type, assuming equal attention to all the feature types while computing the final embedding vector. So, here we assume $\alpha_{ir}=1$ $\forall i,r$.

\begin{table*}[h]
\centering
\scriptsize
\begin{tabular}{llc}
\hline
Feature type & Example features                                                                                                                                                                   & \% drop for one feature type \\ \hline
Hashtags      & \begin{tabular}[c]{@{}l@{}} \#taxreform, \#maga, \#medicareforall,\\ \#healthcare, \#txlege, \#tcot, \#coleg, \#gopwomen\end{tabular}                                       & 2.7\%                         \\ \hline
Domains       & \begin{tabular}[c]{@{}l@{}} senate, clk, fairandsimple,\\ house, theguardian, congress \end{tabular}                                                        & 0.9\%                        \\ \hline

Followees       & \begin{tabular}[c]{@{}l@{}} JimLangevin, txstdems, COHouseDem,\\ Sam1963, MustafaTameez, pdamerica\\  \end{tabular}                                & 0.3\%                         \\ \hline
Retweetees       & \begin{tabular}[c]{@{}l@{}} RepJoseSerrano, PuestoLoco, MustafaTameez,\\ SenatorMenendez, ProjectLincoln, JoeBiden\end{tabular}                     & 0.5\%                         \\ \hline
Repliees       & \begin{tabular}[c]{@{}l@{}} repblumenauer, RepJoseSerrano, SenatorMenendez, Sam1963\end{tabular}                                & 0.3\%                         \\ \hline
Mentions       & \begin{tabular}[c]{@{}l@{}} \@kylegriffin1, \@MustafaTameez, \@WhiteHouse45,\\ \@texasdemocrats, \@JoeBiden, \@Mike\_Pence\end{tabular}                        & 0.5\%                         \\ \hline

\end{tabular}%
\caption{\small{Important features and feature types for the predictions.}}
\label{tab:mostimpfeat}
\vspace{-8mm}
\end{table*}

To test the few-shot learning framework we used incrementally powerful models in Table~\ref{tab:ablation}, where FSSPIP- - - is the base architecture without the few-shot learning framework and then each component of the framework is added sequentially to the base model (terming those intermediate models FSSPIP- -, FSSPIP-, and finally FSSPIP).

We find that the dynamic attention mechanism produces significantly\footnote{We ran a significance test comparing the results of other two variants with the main (dynamic) variant and found the $p$-value to be less than 0.05 in all the cases (compared to both F1 and F2) as marked by \dag~in Table~\ref{tab:ablation}.} higher gains compared to the other two attention variants.
We find that the gains produced are higher when fewer data points are used and weak supervision has a higher impact than adding dynamic augmentation further to the weakly supervised model. This can be explained as weak supervision already trains the model with a large number of real data points which makes the model regularized enough. However, dynamic augmentation helps in regularizing the model, specially in fewshot settings to avoid over-fitting. Similarly, self-supervision also seems more useful in case of fewer training data points. Moreover, we can see that the attention variants of the model perform very close to the original model but falls short with low number of data points.

\noindent\textit{Most important feature types} - To determine the most important feature types, we drop each feature channel and measure the information loss by calculating the deviation in performance of the classifier (FSSPIP) trained on the combined dataset (train:test:validation datapoints = 80:10:10). The results are reported in Table~\ref{tab:mostimpfeat}. The highest drop is witnessed when the relevant hashtags are dropped. 

\noindent\textbf{Zero-shot gain}: Inspired by the significant improvement by weak supervision as shown in Table~\ref{tab:ablation}, we trained our model \textit{FSSPIP} on the weak supervision dataset \textit{only}, which is collected without any manual annotation. We then use the whole annotated dataset for testing this model. We obtain a \textbf{\textit{zero-shot accuracy}} of 93.7\% (TIMME models are based on list of politicians of each party, and thus cannot be zero-shot. UUS, which is not easily scalable due to its clustering, purity checking by experts and soft labelling methodology, performed the best among other baselines at 91.9\%). This tells that the social media followers of a political party are indeed, most of the time, followers of the party in real life also. So, training a model to classify a social media user to be a follower of one party over the other on social media also trains the model for the similar task of classifying the user to be a follower of one political party over the other in real life. We verify this conclusion again in a multi-party scenario for a diverse non-English speaking democracy like India in the next section.\\

\vspace{-5mm}

\section{Additional task: Experiments and analysis}\label{casestudies}

We use the additionally collected datasets to show the efficacy of the zero-shot classifier. The research questions selected for this section are easy to test but important for social scientists. They had been mostly analyzed through manual surveys till now.

\if{0}The research questions we want to address are as follows:
\begin{enumerate}
    \item Can our classifier be applied on related tasks e.g., check the bias of media houses?
    \item Can we identify the leanings for specific issues of importance using the classifier?
    \item Does our classifier work equally well in a multi-party setting?
    \item Can our classifier find out the political leanings of people given the geological localization?
    \item How likely is it for people to change their inclination over time?
    \item Can the classifier find out the political leanings of people tweeting with specific hashtags?
\end{enumerate}\fi

\noindent\textbf{Media bias prediction}: We use the trained FSSPIP classifier on the media bias dataset collected by us, taking each of the Twitter handles of the media houses as the node to be classified. We obtain an accuracy of 72.6\% on the task, while we do not explicitly train for this task\footnote{If we train for the task explicitly using 70:30 split for train-test data, we achieve an accuracy of 84.1\%.} and rely on the assumption that $\{\textrm{democrat}\equiv\textrm{left}\}$ and $\{\textrm{republic}\equiv\textrm{right}\}$. 
\begin{figure}[t]
    \centering
    \begin{subfigure}{0.44\linewidth}
    \includegraphics[width=1.0\textwidth]{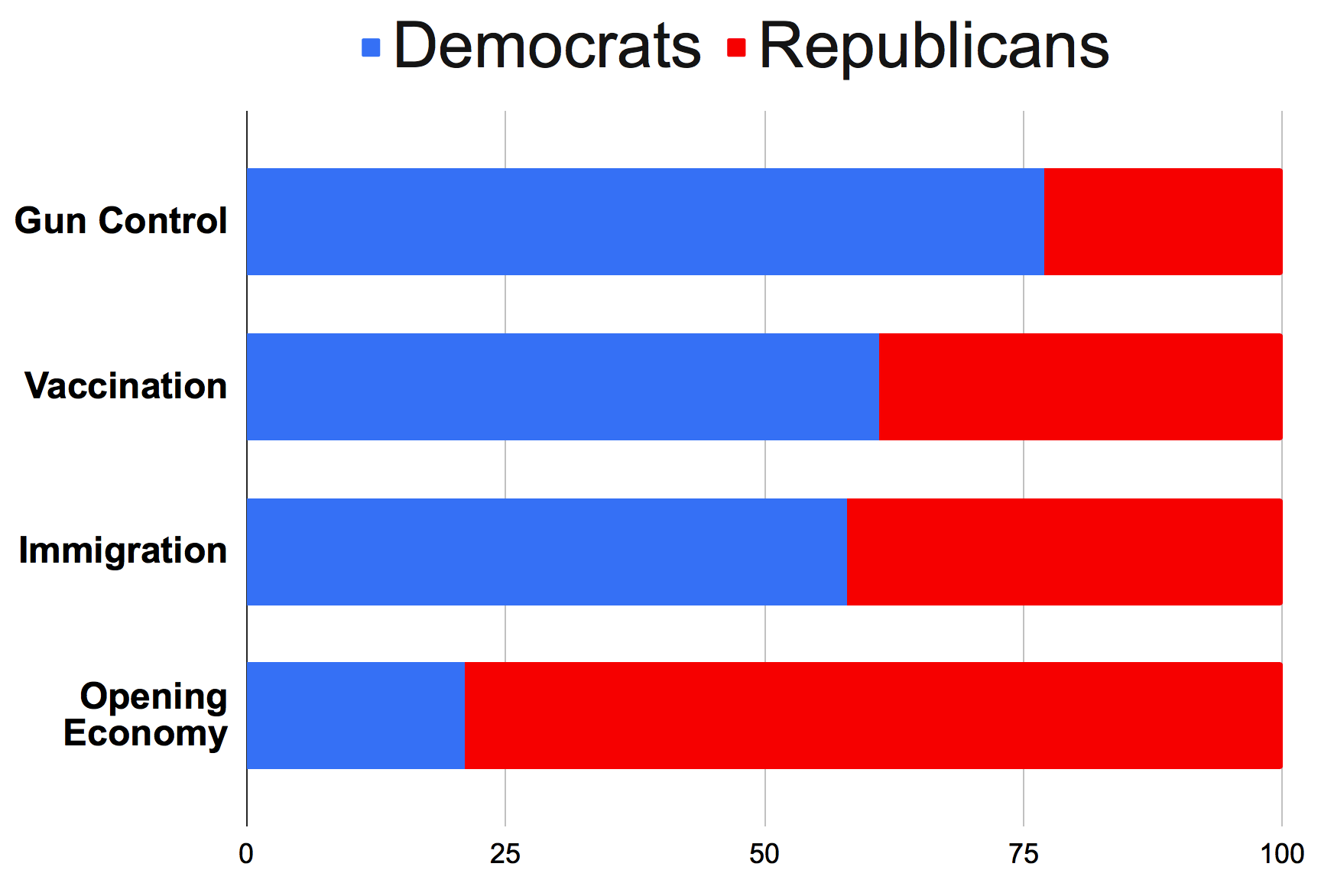}
    \caption{\label{fig:topic}Issue wise political leaning}
    \end{subfigure}
    \begin{subfigure}{0.45\linewidth}
     \includegraphics[width=1.0\textwidth]{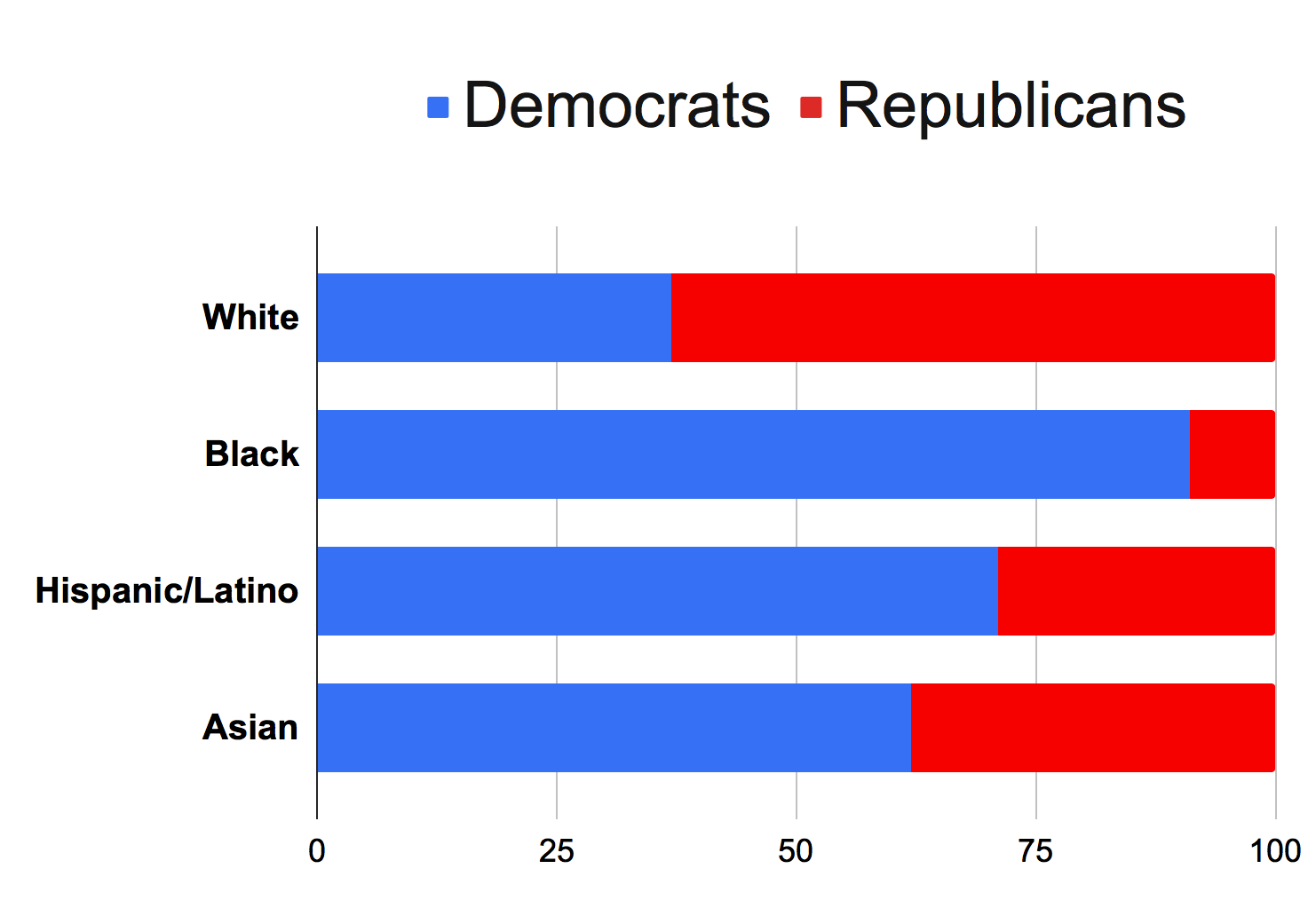}   
     \caption{\label{fig:comm}Community wise political leaning}
    \end{subfigure}
    \caption{\label{fig:commrac} \small{Distribution of political inclinations in the USA by topic/racial demographics.}}
\end{figure}
\if{0}\begin{figure}[t]
    \centering
    \includegraphics[width=0.4\textwidth]{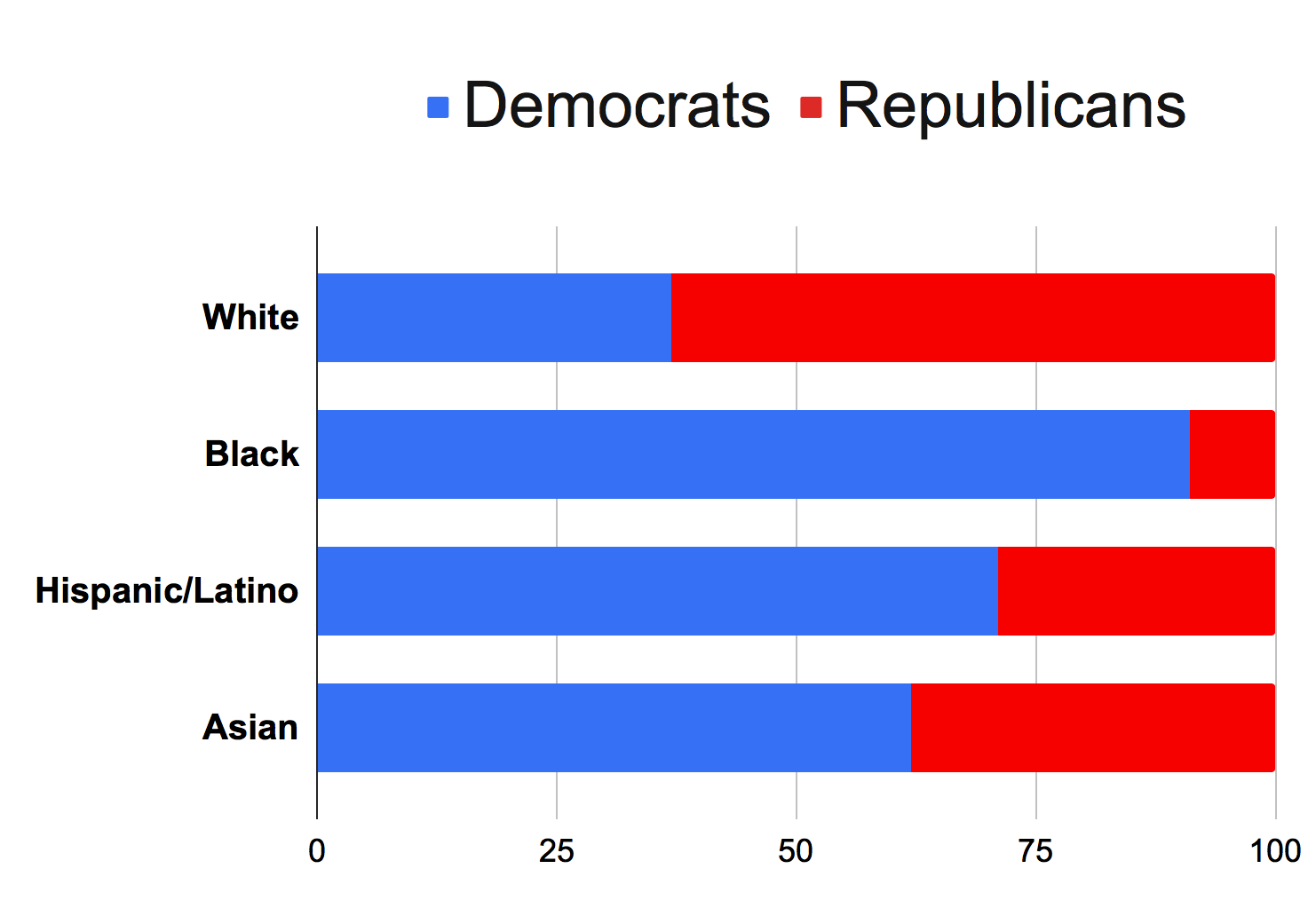}
    \caption{Community wise inclination distribution in the USA.}
    \label{fig:community}
\end{figure}\fi

\noindent\textbf{Topical polarization - bone(s) of contention}: In order to poll users for specific contexts and issues, we collect some hashtags (see Appendix \ref{appendix:hash}) supporting each issue mentioned in Figure~\ref{fig:topic}. We then use the model to classify each user and plot the \% of users for each leaning in the US setting, i.e., \textit{The Democrats} \& \textit{The Republicans}.


\noindent\textbf{Multi-party inclination prediction}: US political system is binary consisting of only two political parties: \textit{The Democrats} \& \textit{The Republicans}.
In principle, our system can work for other countries and other kinds of political systems as well. In this section, we test \textit{zero-shot} property classification of our model on the diverse multi-party democracy like India. We take Twitter handles of three national parties in India, namely, \textit{Aam Aadmi Party} (\textit{AAP}), \textit{Indian National Congress} (\textit{INC}) and \textit{Bharatiya Janata Party} (\textit{BJP}). We use the weak supervision method to train our model with sampling, mixup \& feature channel dropout strategy as discussed earlier. On a random sample of 1000 Twitter accounts (\textit{AAP}: 203, \textit{INC}: 435, \textit{BJP}: 362), we obtain an accuracy of 81.9\%. The highest confusion scores between classes (see Appendix \ref{appendix:confusion}) were between \textit{AAP} \& \textit{INC}. This is fairly intuitive since both these parties are left-leaning and in opposition, while \textit{BJP} is known to be subscribing to a right-wing leaning and is currently in power.

\begin{figure}[t]
\centering
\begin{subfigure}{0.39\linewidth}
    \includegraphics[width=1.0\textwidth]{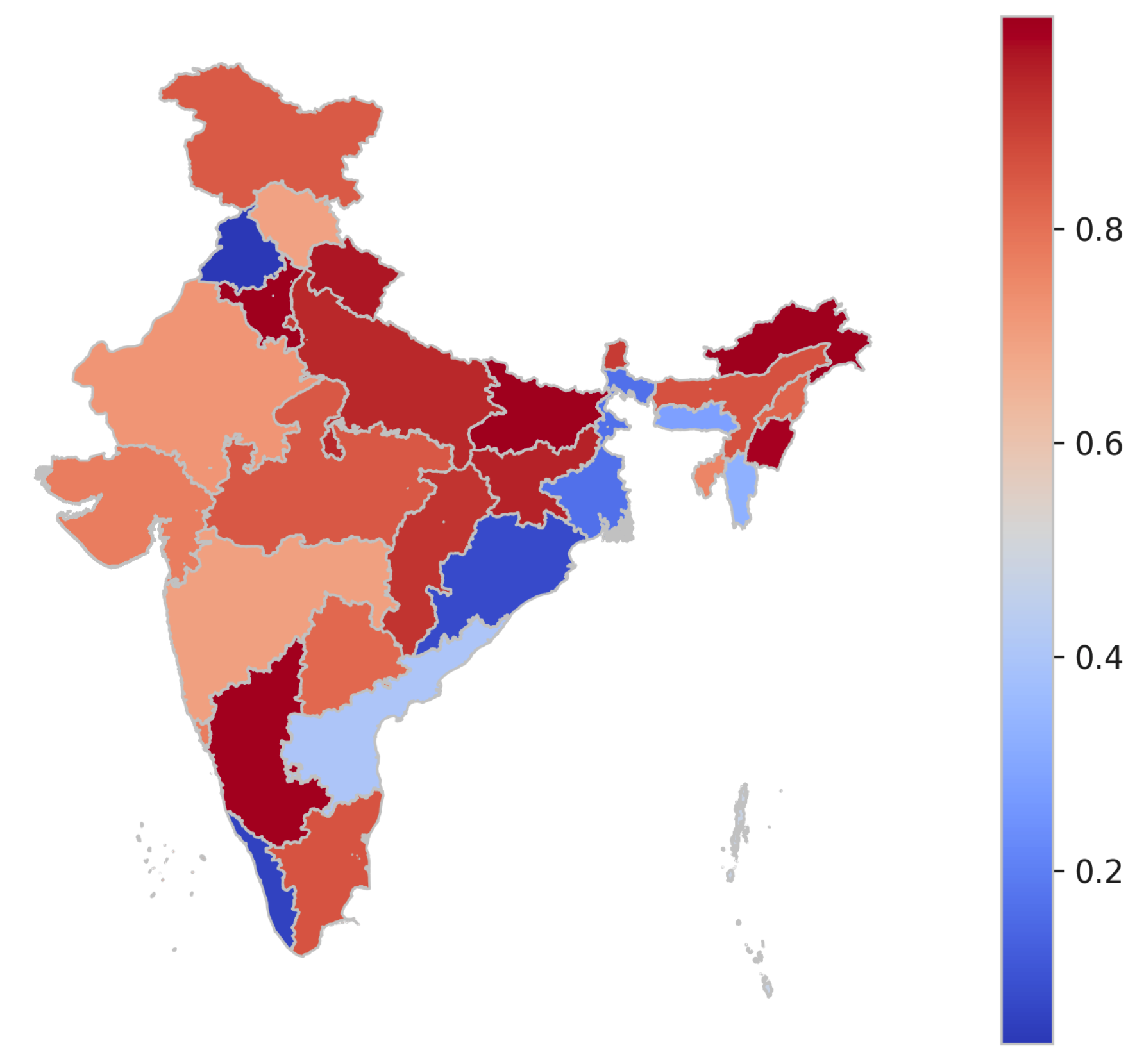}
    \caption{\label{fig:india}Statewise distribution of average political leaning toward \textit{BJP} in India.}
    \end{subfigure}
    \begin{subfigure}{0.55\linewidth}
    \includegraphics[width=1.0\textwidth]{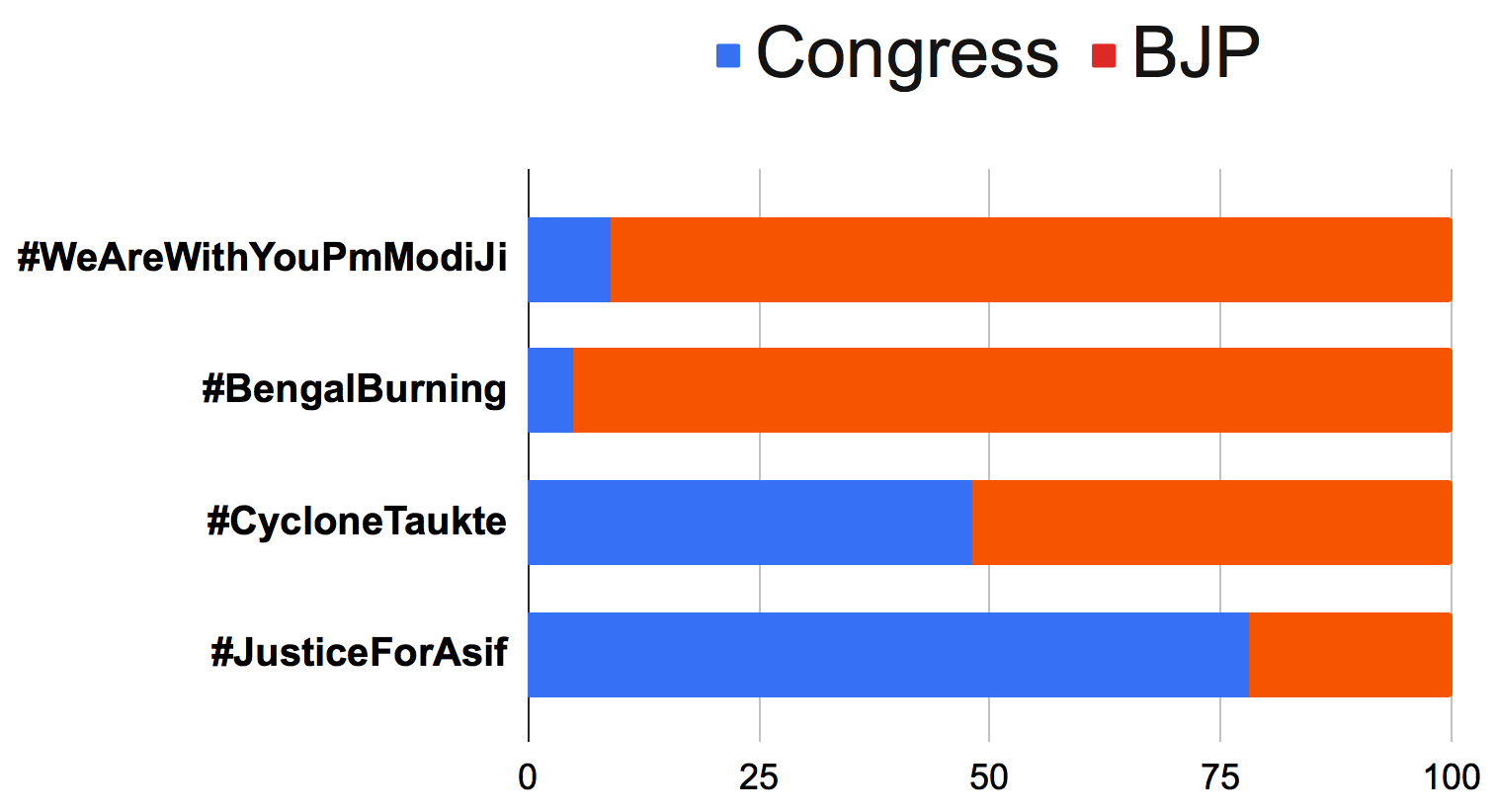}
    \caption{\label{fig:hashtag_india2} Distribution of political leanings of hashtag promoters}
    \end{subfigure}
    \caption{\label{fig:hashtag_india5}\small{Inclination distribution in India.}}
\end{figure}

\noindent\textbf{Statewise leaning}: In Figure~\ref{fig:india}, we plot the relative distribution of political leanings for each state of India (on a scale of 0-1, signifying the percentage of users in a state leaning toward \textit{BJP}. We get the average of political leanings for each person in the state's data predicted using the aforementioned classifier). This correlates quite well (Pearson's  corr coef: 0.52 with high significance and low p-value, $p<0.01$) with the vote percentage received by \textit{BJP} in each state in the 2019 general election.\footnote{\url{https://eci.gov.in/}} 

\noindent\textbf{A leopard cannot change its spots}: In order to check if the political inclination changes with time, we reuse the same dataset described in the last paragraph with a temporal filtering strategy. We only use the tweets and tweet derived features for this experiment which means the bio is always left as blank and same is done for followers/retweeters. We collect all the 3,200 tweets (limit set by the twitter API) of each user ID, directly available from Twitter. For reliable prediction, we filter the users who have tweeted at least 100 times before 2017 and at least 100 times after 2018\footnote{We do not consider the follower-followee network as the past snapshots are not retrievable.} This leaves us with 2,893 users. We then predict the inclination of these Twitter users twice. Once we use the features collected from tweets before 2017 and once we use the tweets after 2018. We observe that the predictions match for 91\% of the cases, which tells that political leanings are temporally (almost) invariant.\\
\noindent\textbf{Hidden agenda - inclination behind promoted hashtags}: To find out the inclination behind each hashtag, we obtain the political leanings of the users in the collected hashtag-specific dataset using the zero-shot classifier trained on followers of Congress and BJP. We plot the percentage of users leaning toward each party for each hashtag. We correctly predicted the leaning in 25 of the 30 cases using the classifier (considering a percentage distribution of 40-60\% as the neutral/apolitical zone). We plot the leanings on four different India-specific issue -- \textit{\#WeAreWithYouPmModiJi}, \textit{\#BengalBurning}, \textit{\#CycloneTaukte}, \textit{\#JusticeForAsif} in Figure~\ref{fig:hashtag_india2}. While we see the disaster hashtag (\#CycloneTaukte) is non-polarizing, other trending hashtags are evidently promoted by people of particular ideologies. We include the list of other hashtags in Appendix \ref{appendix:hashindia}.

\if{0}\begin{figure}[t]
    \centering
    \includegraphics[width=0.4\textwidth]{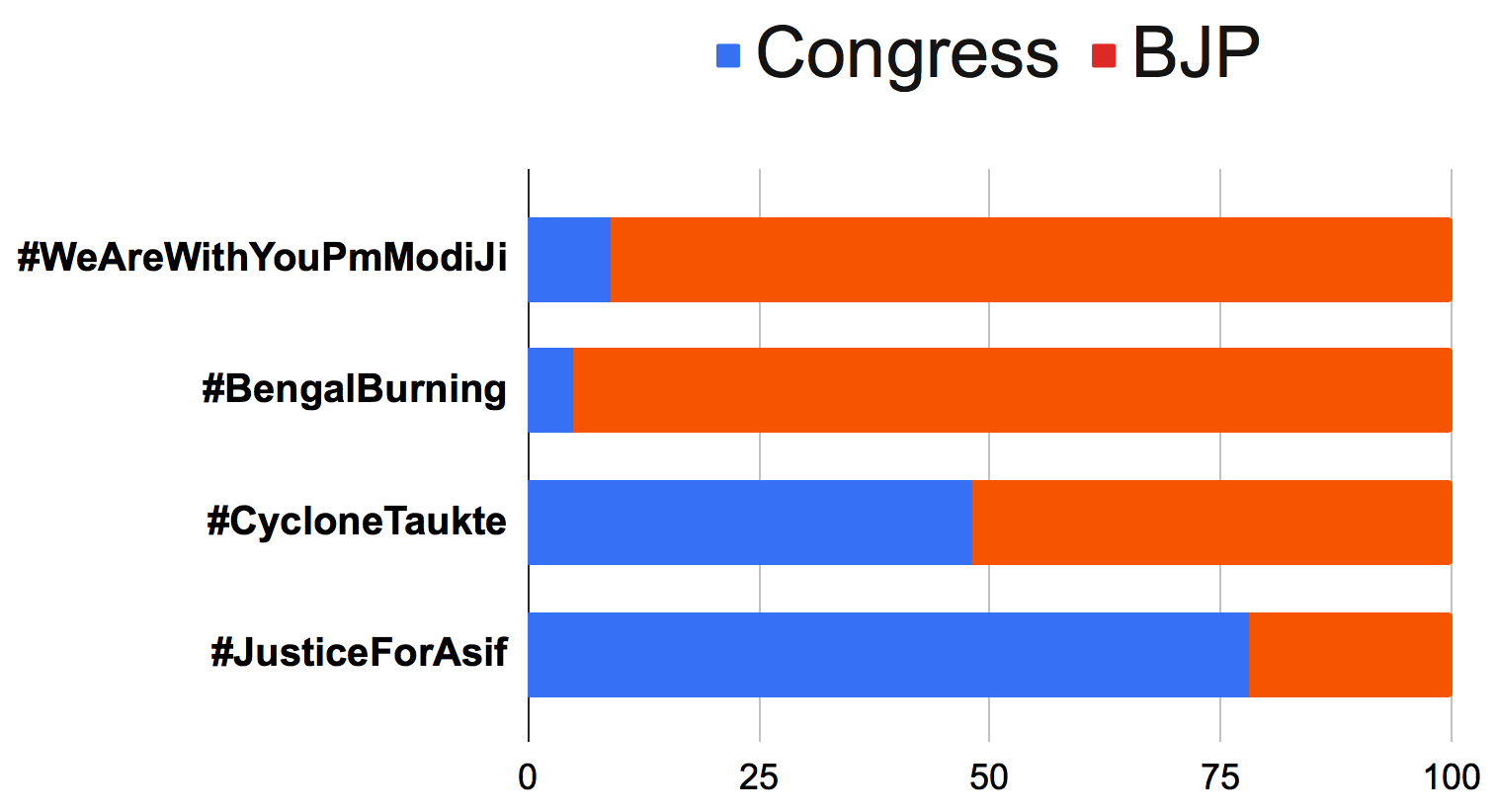}

\end{figure}\fi
\vspace{-0.2cm}
\section{Limitations and future work}
\vspace{-0.2cm}
Our work is limited by the availability of social media data. If a country does not have enough political participation in the social media, then training a model will not be possible. Moreover, if the profile of a person is kept private, the classifier will not be able to assign any label. We have discussed the related ethical implications of our work seperately in Appendix~\ref{ethics} 

Lastly, here we only evaluated our method on a dataset of users with high degree of political connection to very low degree of political connection. Collection of a dataset of users with \textit{no political links} online but inclination toward a particular political party is a challenging task. In fact, Twitter matched voter registration data~\cite{barbera2015birds} also shows high partisanship evident in tweets and political connections. Research toward implicit(not explicitly tweeted/mentioned) political inclination detection (like implicit hate speech detection
~\cite{latent2021,scrush} or implicit aspect specific sentiment detection~\cite{implicit_senti,saspect}) is an interesting future research direction.

\vspace{-0.4cm}
\section{Conclusions}
\vspace{-0.4cm}

We present an efficient, fast, and scalable few-shot learning framework for Twitter profiles for political inclination detection (FSSPIP). We showed that our model is explainable and learns features that humans find meaningful. Moreover, our model does not store any personal data of users unlike graph based models. 
It is also shown to be faster than graph-based methods. With the scalable representation learning framework, we achieve state-of-the-art accuracy, 
gaining significantly in  unlabelled or few-shot learning setups on non-politician users. Enabling \textit{zero-shot political inclination detection} with high fidelity, we provide a method to easily re-target this work to new countries and languages without any manual intervention/supervision unlike previous methods. We believe this will make a large-scale analysis of the political landscape throughout the globe easier and more accurate. We intend to make the codebase for data collection and prediction available once the paper is accepted.

\bibliography{main}
\bibliographystyle{splncs04}

\newpage

\onecolumn
\appendix
\section{Feature processing}\label{appendix:a}
We use a total of 22 feature types. Specifically for each user the core feature types are:
\begin{enumerate}
    \item Text 
    \item Hashtags
    \item Domain names
    \item Domain+Co-domain names
    \item Mentions
    \item Bios
    
\end{enumerate}
We extract the above mentioned features from the tweets of the users.

We use these same set of core features with replies:
\begin{enumerate}
    \item UserID Replied to
    \item Hashtags used in replies
    \item Domain names used in replies
    \item Domain+Co-domain names used in replies
    \item Mentions in replies
    \item Bios of the UserID Replied to
    
\end{enumerate}
For quote tweets and retweets:
\begin{enumerate}
    \item UserIDs Retweeted
    \item Hashtags in original tweets
    \item Domain names in original tweets
    \item Domain+Co-domain names in original tweets
    \item Mentions in original tweets
    \item Bios of the UserIDs Retweeted
    
\end{enumerate}

Finally, userIDs of followers and friends, ids of users retweeted and replied to are taken as features making a total of 22 feature types.

\section{Total time calculation}
In order to compute total time we take the userIDs as the inputs whose ideologies are to be detected. The whole pipeline is then run from user data collection to inference. Average of runs for all users have been reported as the total time to inference (TTI).
\begin{figure}[ht]
    \centering
    \includegraphics[width=0.45\textwidth]{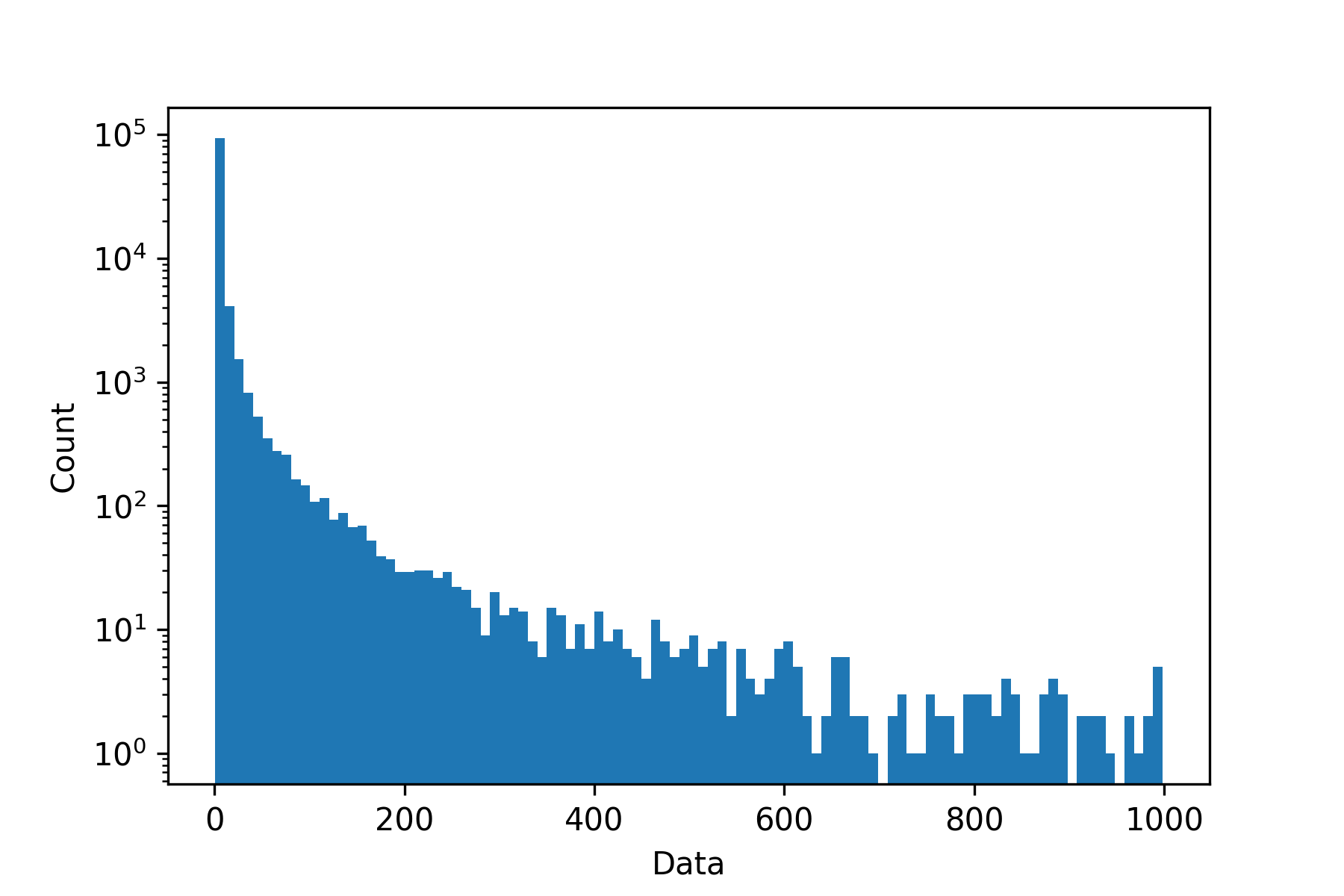}
    \caption{Frequency Distribution of hashtags, for the most frequently mentioned 1000 hashtags in the \textbf{P+all} dataset}
    \label{fig:hashtags}
\end{figure}
\begin{figure}[ht]
    \centering
    \includegraphics[width=0.45\textwidth]{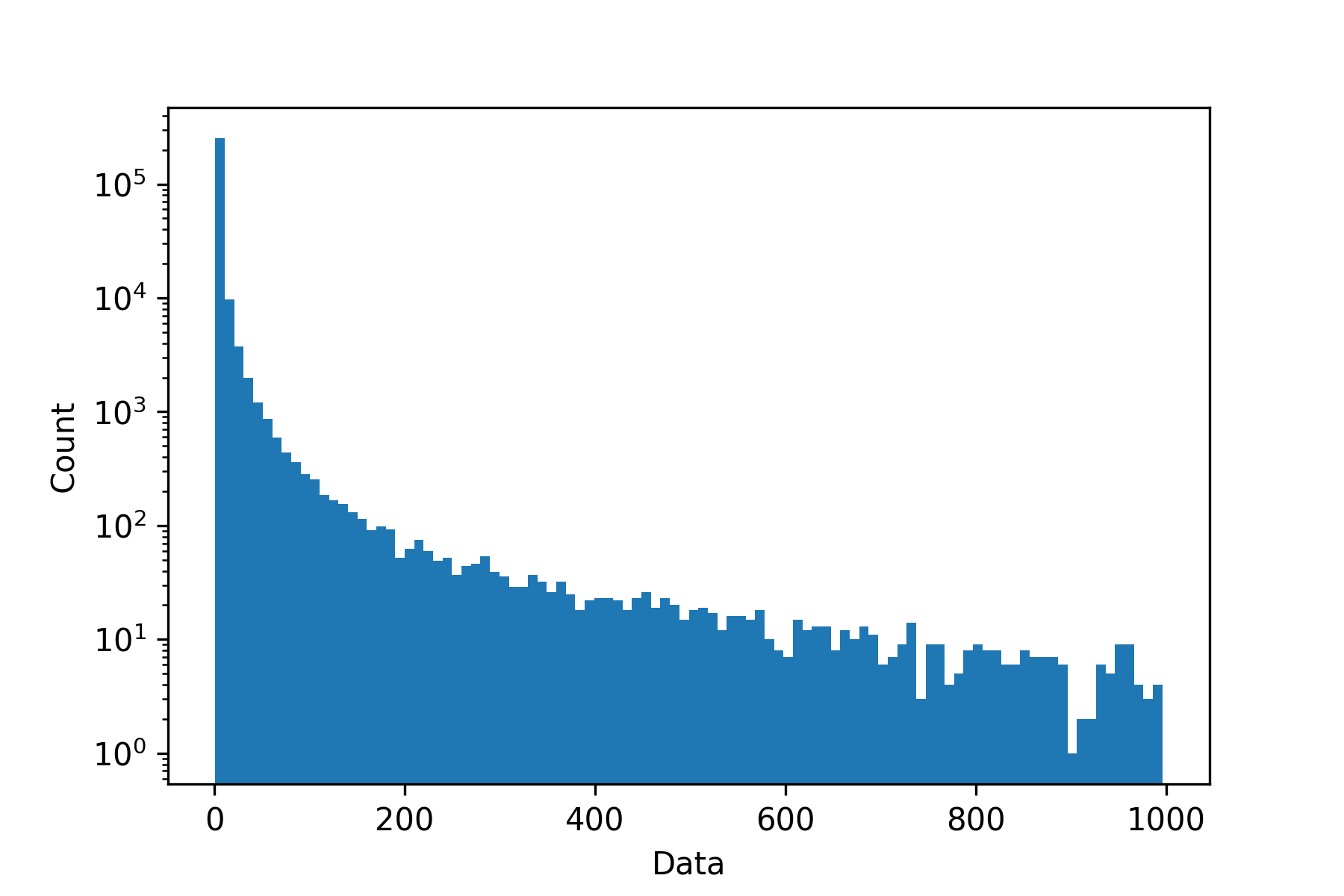}
    \caption{Frequency Distribution of mentions of userIDs, for the most frequently mentioned 1000 userIDs in the \textbf{P+all} dataset}
    \label{fig:users}
\end{figure}
\section{Data distribution}

In the section on feature processing, we have described the number of unique features for the different feature types. However, the frequency of these features are not uniformly distributed.
We plot the distribution of two top features: \textit{hashtags} (Fig~\ref{fig:hashtags}) and \textit{mentions} (Fig~\ref{fig:users})  for the \textbf{P+all} dataset (which is the closest to a real world distribution of users) for a better visualization and understanding of the dataset.
We see from the figures that the features mostly follow a exponentially decreasing frequency curve meaning that a handful of hashtags and user mentions mostly dominated the Twitter space of the users in our dataset which is very intuitive for any real world social media dataset.

\section{Hyperparameters} 
We mention the number of instances used in Table 3 in the main paper. We use 10\% data for testing and 10\% data for validation using 80\% data for training. We compute the average of five runs for each model for all the reported numbers. For the non-neural baselines, we use the scikit-learn library. We use a maximum of 30 epochs till convergence on the validation set. For our model, we use the PyTorch library\footnote{\url{https://pytorch.org/}} with a batch size of 32, a learning rate of 0.01, and Adam optimizer. We use the default library-defined hyperparameter values for all other cases. We trained all our models (\textit{FSSPIP variants}) for 50 epochs. 
We ran our experiments on Tesla P100 16GB GPUs. It took less than an hour in training time to train the classifier.

\section{Additional Equations for the Fewshot Learning Framework}
\noindent{\textbf{Modified MixUp for our case}}

\begin{equation}
    \tilde f_i= sample_p(f1_i)+sample_{1-p}(f2_i) 
\end{equation}
when $p \in \beta(\alpha,\alpha)$ and $\alpha=0.1$, p is the sampling probability of choosing a specific feature if it is present in that feature channel, $f_i$ is the modified feature vector for the i\textsuperscript{th} feature channel for the augmented user datapoint $\tilde u$ and $fx_i$ is similarly the i\textsuperscript{th} feature channel vector for the x\textsuperscript{th} user $u_x$. Here $u_1$ and $u_2$ are two randomly sampled users from the datapoints.

\noindent{\textbf{Loss of Self-supervision}}

Let the l\textsuperscript{th} user be $u_l$ who is denoted by a list of one hot feature vectors $\{f_{l1},f_{l2},f_{l3},..f_{ln}\}$ where $f_{lij}\in f_{li}$ is the j\textsuperscript{th} feature.
Now, if the j\textsuperscript{th} feature is masked randomly, we try to predict it using the cross-entropy loss.

The self-supervised loss would be calculated as:
\begin{align}
    L_{\text{ss}} = \E_{l \sim \mathcal{U}([N])} \left[ \sum_i \sum_j - log \frac{\textmyfont{exp}(F[f_{lij}])}{\sum_{k\in[\mathcal{K}_i]}\textmyfont{exp}(F[f_{lik}])} \right]
\end{align}

where $\mathcal{U}([N])$ is the uniform distribution on N users in the training set and F is the feed forward network on top of the base FSSPIP architecture's final encoding layer that estimates the probability of presence of the masked feature $f_{lij}$ and $\mathcal{K}_i$ is the list of features in feature channel $\mathcal{K}_i$.
\noindent{\textbf{Hyperparameters of Self-supervision}}

We pretrained the model on two 16GB Tesla P100 GPUs. We used a learning rate of 3e-5 and trained the model for 5 epochs on the unannotated dataset. For other hyperparameters we used the defaults in the Pytorch library.

\section{Detailed result on whole training data}
\label{appendix:performance}
In Table~\ref{tab:perfstat}, we have put the results when we use the whole training data instead of the few-shot settings. We can see that the FSSPIP variants strike a perfect balance between performance, speed and user privacy. While the FSSPIP variants perform significantly better on the testset than the non-DL baselines, it is atleast \textbf{10x faster} than \textit{the best-performing DL baseline requiring no human intervention} and more scalable than the UUS+ model which requires political experts' intervention while offering comparable performance and respecting user privacy as mentioned in section \ref{ethics}.

\noindent{\textbf{Total Loss for pretraining}}
So the total loss used for pretraining is:
\begin{equation}
    L_{pretraining}=L_{mixup} +L_{ss}
\end{equation}
where $L_{mixup}$ is the vanilla cross-entropy loss used on the augmented datapoints created by the MixUp technique
\begin{table}[t]
\centering
\scriptsize 
\begin{tabular}{l|ll|ll|ll|ll|l}
\hline
\multirow{2}{*}{Model} & \multicolumn{2}{l|}{PureP} & \multicolumn{2}{l|}{P50} & \multicolumn{2}{l|}{P20-50} & \multicolumn{2}{l|}{P+all} & \multirow{2}{*}{TTI (secs)} \\ \cline{2-9}
                       & Acc          & F1          & Acc         & F1         & Acc          & F1           & Acc          & F1                                                &                            \\ \hline
Majority               & 53.2         & 69.4          & 54.9        & 70.9         & 61.8         & 76.4           & 60.6         & 75.5                                             & - -                         \\ \hline
LR                     & 90.2         & 90.3        & 82.5        & 82.4       & 94.3         & 93.3         & 89.2         & 88.7                                        & 27.96                      \\ \hline
RF                     & 98.2         & 98.2        & 87.1        & 84.7       & 90.2         & 90.3         & 85.3         & 84.9                                       & 28.9                      \\ \hline
SVM                    & 92.2         & 92.6        & 70.6        & 67.8       & 89.9         & 88.2         & 85.2         & 84.2                                        & 28.1                      \\ \hline

NTF                    & 96.9         & 96.5        & 95.1        & 95.2       & 95.8         & 95.7         & 95.1         & 95.2                                        & 65.3                      \\ \hline

UUS                    & 94.1        & 93.9        & 92.2        & 92.4       & 92.5         & 92.5         & 89.1         & 89.2                                        & 63.1                      \\ \hline

UUS+                    & 95.2         & 95.3        & 95.4        & 95.6       & 95.0         & 94.8         & 93.9         & 94.1                                        & 63.2                      \\ \hline

TIMME-single          & \textbf{99.5}         & \textbf{99.6}        & 97.4        & 97.8       & 98.3         & 98.2         & 93.2         & 93.5                                          & 970.1                         \\ \hline
TIMME            & 98.2         & 98.3        & 98.4        & 98.1       & 99.2         & 99.3         & 95.6         & 95.3                                     & 971.5                         \\ \hline
TIMME-hier             & 99.2         & 99.3        & 97.7        & 97.2       & \textbf{99.8}         & \textbf{99.5}         & 96.8         & 96.7                                       & 973.2                         \\ \hline
FSSPIP-auto        & 96.2         & 96.9        & 95.9        & 95.6       & 96.1         & 96.7         & 94.3         & 94.9                                         & 61.2                         \\ \hline
FSSPIP-fixedattn           & 97.5         & 97.8        & 96.3        & 96.2       & 98.1         & 98.7         & 97.3         & 97.8                                         & 61.8                         \\ \hline
FSSPIP-dyattn             & 99.1         & 99.3        & \textbf{99.1}        & \textbf{98.9}       & 98.9         & 98.8         & \textbf{98.3}         & \textbf{97.9}                                      & 62.1                         \\ \hline
\end{tabular}%
\caption{Performance of different models on the whole training data in terms of accuracy, F1-score and total time to inference (TTI) per user.}
\label{tab:perfstat}
\end{table}

\section{Ethical considerations}\label{ethics}

This model is meant to be used by social scientists in settings similar to the case studies shown by us (sometimes with other demographic classifiers~\cite{sdecoding,sgaurav}). This model as shown in Section~\ref{casestudies}, can also be used by media bias researchers to understand the political biases~\cite{smediabias} in media houses and the audiences of specific organizations or influencers in social media. Moreover, this needs to be done in a privacy-preserving zero-shot setting, as used in this paper.

However, there are wider issues related to the potential of such a tool to be abused in the hands of malicious entities. Knowing the political orientation of a large number of users could potentially expose them to harm in case of a sudden regime change. So, such tools should be used with caution and any service using such a tool must be allowed to only report aggregate statistics for journalistic use and no personally identifiable data should be reported.

\section{List of hashtags in US setting topicwise}
\label{appendix:hash}

Below is the list of hashtags used for each topic in US setting for the experiments in the \textit{\textbf{Topical polarization}} section of the main text.

\noindent{\textbf{Gun control:}} \#Guncontrol, \#Gunviolence, \#guncontrolnow, \#endgunviolence , \#gunsafety, \#gunlaws, \#gunsense

\noindent{\textbf{Vaccination:}} \#vaccinationday, \#babyvaccination, \#immunizationday, \#provaccination

\noindent{\textbf{Immigration:}} \#immigrantsmakeamericagreat,  \#permanentresidency, \#immigrationusa, \#immigrationrights

\noindent{\textbf{Opening the economy:}} \#takeamericabackagain, \#endthelockdown, \#OpenAmerica, \#TakeAmericaBack, \#NoLockdown

\section{Confusion matrix for India\\ specific multiparty classification}\label{appendix:confusion}
\begin{figure}[h]
    \centering
    \includegraphics[width=0.4\textwidth]{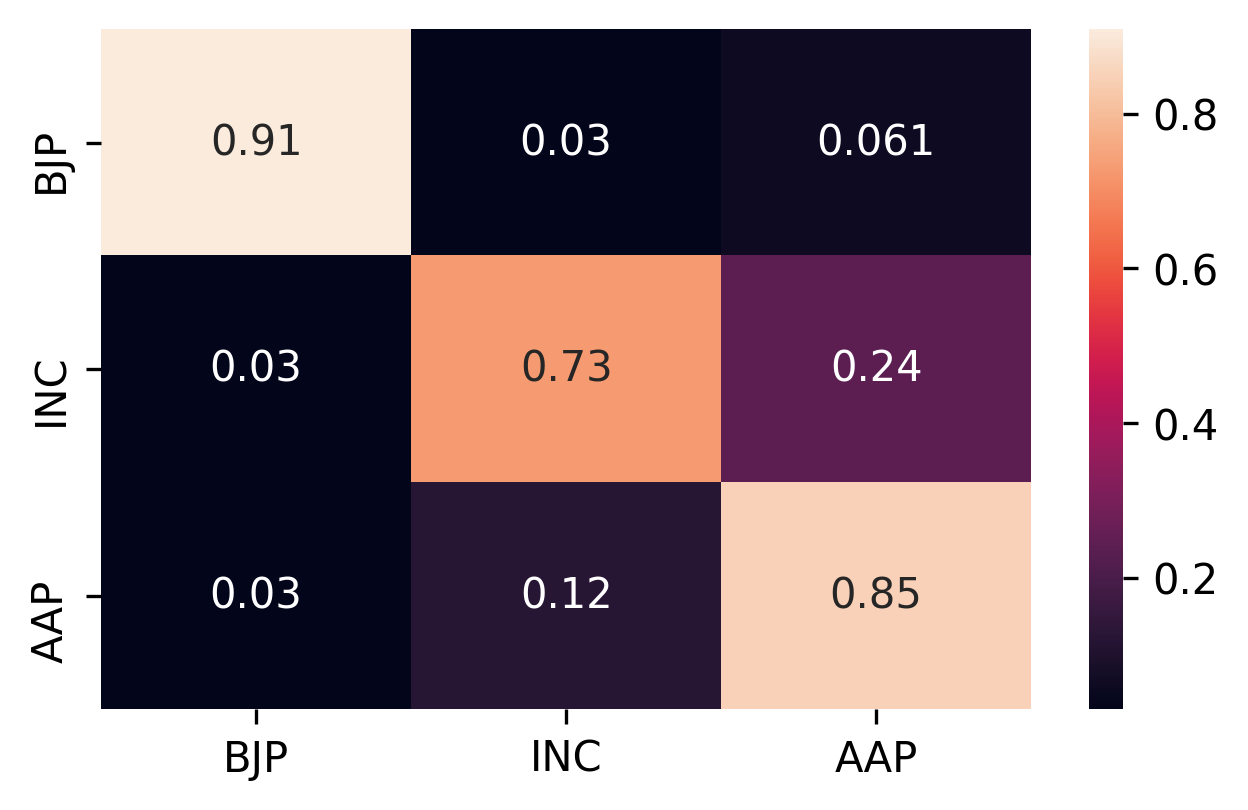}
    \caption{Confusion matrix: Zeroshot classification of 1000 Indian twitter handles for three classes(three political parties)}
    \label{fig:fig18}
\end{figure}

Figure~\ref{fig:fig18} shows the confusion matrix in the 3-class classification. We can see that the classifier is particularly confused about AAP and Congress party.

\section{List of hashtags for UUS/UUS+} \label{appendix:new}
\noindent{\textbf{Republican Hashtags: }}\#MAGA, \#MAKEAMERICAGREATAGAIN, \#NeverHillary

\noindent{\textbf{Democrat Hashtags: }}\#ImWithHer, \#VoteBlue, \#VoteHimOut

\section{List of 30 hashtags in Indian setting}
\label{appendix:hashindia}
\noindent{\textbf{BJP leaning hashtags}}
\begin{verbatim}
#WeAreWithYouPmModiJi
#NATION_WITH_MODI
#BengalBurning
#andolonjeevigaddarhay (hindi)
#Bhrast_Congress (hindi)
#HinduVirodhiCongress
#postpollviolence
#bengalviolence
#HindusLivesMatter
#ModiHaiTohMumkinHai
\end{verbatim}

\noindent{\textbf{Neutral hashtags}}
\begin{verbatim}
#CycloneTaukte
#WorldHypertensionDay
#modiji_cancel12thboards
#MumbaiRains
#BAPSSwaminarayanFraud
#mondaythoughts
#MondayMotivation 
#LifeSaverRajeshwarHospital
#BitCoin
#BoycottIsraeliProducts
#UWCL
#WeWantChinmayiBack
\end{verbatim}

\noindent{\textbf{Congress leaning hashtags}}
\begin{verbatim}
#JusticeForAsif
#JusticeForAsifa
#MobLynching
#sada_huya_samaj (hindi)
#khattar_sarkar_gunda_sarkar(hindi)
#BengalModel
#KhelaHobe
#Islamophobia
\end{verbatim}

\end{document}